\documentclass[aps,prl,twocolumn,superscriptaddress,showpacs]{revtex4}
\def\av#1{\left\langle#1\right\rangle}

\def\di{\partial}
\usepackage{amstext,amssymb,amsfonts,amsmath,eucal,graphicx,bm}
\usepackage{epsfig}

\begin{document}

\title{Epidemic threshold for the SIS model on random networks}

\author{Roni Parshani}
\author{Shai Carmi}
\author{Shlomo Havlin}
\affiliation{Minerva Center \& Department of Physics, Bar-Ilan
University, Ramat Gan, Israel}

\date{\today}

\begin{abstract}
We derive an analytical expression for the critical infection rate
$r_c$ of the susceptible-infectious-susceptible (SIS) disease spreading model on random networks. 
To obtain $r_c$, we first calculate the probability of reinfection, $\pi$, defined as the
probability of a node to reinfect the node that had earlier infected
it. We then derive $r_c$ from $\pi$ using percolation theory. We
show that $\pi$ is governed by two effects: (\emph{i}) The
requirement from an infecting node to recover prior to its
reinfection, which depends on the disease spreading parameters; and
(\emph{ii}) The competition between nodes that simultaneously try to
reinfect the same ancestor, which depends on the network
topology.

\end{abstract}
\maketitle

Diseases spread in a population as infected individuals infect
other individuals with whom they are in contact. In recent years,
several models have been developed to mathematically characterize
the spread of diseases \cite{Non_network_epidemics1,Non_network_epidemics2,Non_network_epidemics3}.
The disease spreading is best modeled as a sparse network, where
the individuals are the nodes and the contacts are the links
connecting them
\cite{Network_epidemics1,Network_epidemics2,Network_epidemics3}.
Two epidemic models of particular importance are the SIR
(susceptible/infectious/recovered) \cite{SIR} and SIS
(susceptible/infectious/susceptible) \cite{SIS,SIS2,SIS3}.
In these models, the $N$ network nodes are initially
susceptible (they currently do not have the disease, but might become infected later), except for one randomly chosen infected node. At each time step, an
infected node infects each of its neighbors with probability $r$.
Infected nodes remain as such for $\tau$ time steps, after which
they become either recovered (cannot further infect or become
infected) in SIR or susceptible again in SIS. Hence, in SIS
individuals can be infected multiple times.

A fundamental question in the study of epidemics is: will a disease
spread throughout the population, or will it die out? The answer to
this question depends on the values of the infection and recovery
rates, as well as on the nature of the connections between the
individuals. For the SIR model on random networks, the epidemic
threshold and the critical infection rate above which the disease
infects a non-zero fraction of the population were previously derived
\cite{Network_epidemics3,Percolation_SIR1,Percolation_SIR2}. However, for the
SIS model, the calculation of the critical infection rate is
significantly more involved due to the possibility of multiple
infections of the same node.

Analytical results for SIS on networks exist for a number of
specific network models (e.g., a circle, a lattice, a double-layered
fully connected network, etc.
\cite{Liggett,Diekmann98,Ball99,Neal08}). A mean-field analysis of
SIS on general random networks has been proposed in
\cite{PS,Boguna}. In this approach, master equations are written
for the number of infected individuals of a given degree. To
calculate the epidemic threshold, the mean-field approach assumes
that an infected individual has an equal probability to infect each of its neighbors. 
However, as we show below, the probability of an individual to
reinfect the neighbor that had previously infected it (the
reinfection probability) is different from the probability to infect
the other neighbors.

In this Letter, we derive an analytical expression for the SIS epidemic threshold.
We find that the critical infection rate at the threshold is higher than the mean-field approximation for SIS but lower
than the critical infection rate for SIR. Simulation results support the accuracy of our expression.
To obtain the epidemic threshold, we first calculate the reinfection probability which depends on both the disease spreading
parameters and the network topology. Then, using percolation arguments we derive the threshold from the reinfection probability.

To derive the epidemic threshold, some definitions from percolation theory are
needed. In a percolation process on a network, links are removed
until only a fraction $p$ of the $N$ network nodes remain. This is
continued until a critical value, the percolation threshold $p_c$,
is reached. For $p>p_c$, a spanning cluster of order $N$ nodes
exists, while for $p<p_c$ the network collapses into small clusters
\cite{Percolation,Percolation3,ER,Bollobas,Cohen}.

\begin{figure}
\begin{center}
\epsfig{file=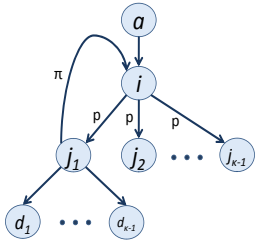,height=4cm,width=4cm}
\end{center}
\caption{A schematic illustration of the sub-network
in which the probabilities of infections are calculated in the text.}
\label{schematic}
\end{figure}

Disease spreading can be seen as a growing percolation process \cite{Leath},
in which, starting from a given seed, links are being added to the
growing network with probability $p$. The critical infection rate in
which the disease spreads throughout the network is
equivalent to the percolation threshold in which a spanning cluster
appears. To complete the mapping, we specify the epidemic equivalent
of $p$, that is, the probability of a node to infect its neighbor.
This probability is different from $r$, since an infected node $i$
(see Fig. \ref{schematic}) can infect its neighbor $j$ only as long
as it is still infected, for $\tau$ time steps. Therefore, the
desired probability is given by \cite{Percolation_SIR2}:
\begin{equation}
\label{p_definition} p=1-(1-r)^{\tau}.
\end{equation}
The critical infection rate $r_c$ for a given $\tau$ can then be obtained by
substituting $p_c$ in Eq. \eqref{p_definition}.

To find $p_c$, we define $\av{n_i}$, the average number of
susceptible nodes infected by an already infected node $i$. If
$\av{n_i}>1$, the disease will keep on spreading until a non-zero fraction of the
network is covered \cite{condition_epidemics}. In SIR, only the
descendants of a node should be taken into account, whereas in SIS,
the ancestor could also be reinfected. In terms of Fig.
\ref{schematic}, node $i$, once infected, can infect not only its
descendants $j_1,j_2,...$ (in SIR) but also its ancestor $a$ (in
SIS).

The probability of a node that is reached by following a link to have degree
$k$ (one incoming link and $k-1$ outgoing links) is $kP(k)/\av{k}$ where P(k) is the degree distribution of the network nodes.
Therefore, as long as the network has a tree like structure and loops are
negligible \cite{Cohen}, the average number of neighbors infected by
node $i$ for SIR is given by \cite{condition_epidemics}:
\begin{equation}
\label{ni_sir} \av{n_i}_{\textrm{SIR}} = p\sum_k
\frac{kP(k)}{\av{k}}(k-1) \equiv p(\kappa-1),
\end{equation}
where $\kappa-1\equiv [\av{k^2}/\av{k}]-1$ is the branching factor. At
the epidemic threshold, $p_c(\kappa-1)=1$ \cite{Cohen}. Writing
$p_c$ in terms of $r_c$ and $\tau$ (Eq. \eqref{p_definition}) we
obtain:
\begin{equation}
r_c = 1 - \left(\frac{\kappa - 2}{\kappa - 1}\right)^{1/\tau}.
\label{rc_sir}
\end{equation}

In SIS, the additional link leading to the ancestor should be
incorporated into the calculation. This can be accomplished
by making use of an effective branching factor, obtained by
replacing $\kappa-1$ by $\kappa$ (the mean field approach). For
example, consider node $j_1$ in Fig. \ref{schematic}: in addition to
the $\kappa-1$ links that are connected to the descendants
$d_1,...,d_{\kappa-1}$, it also has the link to $i$.

However, this approach is too simplistic: the probability of $j_1$
to reinfect its ancestor is different from its probability to infect
its descendants. This is true since as opposed to the descendants,
$i$ is not necessarily susceptible. This is a combined outcome of
two effects. First, since $i$ must have been sick before infecting
$j_1$ in the first place, $j_1$ can only reinfect $i$ after $i$ has
recovered and became susceptible again (the ``time'' factor).
Moreover, one of the \emph{other} descendants of $i$
($j_2,...,j_{\kappa-1}$ in Fig. \ref{schematic}) might reinfect $i$
before $j_1$ does (the ``neighbors'' factor). Below, we derive an
analytical expression for the probability of reinfection $\pi$,
taking into account these effects and comparing it to the mean field
approach.

Using $\pi$, an expression for $\av{n_i}_{\textrm{SIS}}$ immediately
follows:
\begin{equation}
\label{ni_sis} \av{n_i}_{\textrm{SIS}} = p(\kappa-1)+\pi.
\end{equation}
The first term is the SIR branching factor (Eq. \eqref{ni_sir}) and
the second is the SIS-specific contribution of the probability to
reinfect the ancestor.

Our analysis depends on the assumption that
the descendants of an infected node $d_1,...,d_{\kappa-1}$ are
susceptible and therefore the probability to infect them is $p$.
This assumption is commonly used in epidemic models when studying
disease spreading (e.g.,
\cite{Percolation_SIR2,Network_epidemics3}), but limits the validity
of the results to the region of at or below the epidemic threshold.
For SIR this assumption is trivially true since reinfections
are not permitted and the disease spreads directionally down a tree
structure. However, for SIS where reinfections are allowed, the disease can
spread down and back up the same branch, and thus the neighbors of a
just infected node may already be infected. Nevertheless, since at
or below the threshold the infection rate is already low, the
probability that the disease has spread down and then back up is
extremely small \cite{Note4}.

To calculate $\pi$ we first obtain the probability $\pi_t$ of $j_1$ to infect $i$,
considering the ``time'' factor and ignoring the ``neighbors'' factor. Assume node $i$ has been infected
for $s$ time steps before infecting $j_1$. Since the total lifetime
of the disease is $\tau$ time steps, $i$ remains infected for
$(\tau-s)$ time steps after infecting $j_1$. Therefore, the total
time in which $j_1$ is infected and $i$ is not is $\tau-(\tau-s)=s$.
The desired probability $\pi_t$ is obtained by conditioning on $s$:
\begin{equation}
\label{pi_t}
\pi_t = \sum_{s=1}^{\tau}\frac{(1-r)^{s-1}r}{1-(1-r)^{\tau}}[1-(1-r)^s] = \frac{1-(1-r)^{\tau+1}}{2-r}
\end{equation}
where $(1-r)^{s-1}r/[1-(1-r)^{\tau}]$ is the probability that $i$
infected $j_1$ at step $s$ given that $j_1$ was eventually infected,
and $[1-(1-r)^s]$ is the probability that $j_1$ infected $i$ in at
most $s$ steps.

To complete the derivation of  $\pi$, we incorporate the
effect of competition (which node is the first to reinfect i) between $j_1$ and the other descendants of
$i$. Since $i$ is arrived at by following a link from $a$, it will
have on average $\kappa-1$ descendants, $j_1,...,j_{\kappa-1}$. In
addition, $i$ can also be reinfected by its ancestor $a$. Therefore
\cite{Note1}:
\begin{equation}
\label{formula_pi} \pi = \pi_t \sum_{k'=0}^{\kappa-1}
\binom{\kappa-1}{k'}\frac{\left(p
\pi_t\right)^{k'}\left(1-p\pi_t\right)^{\kappa-1-k'}}{k'+1}.
\end{equation}
To see this, recall that $\pi_t$ is the probability of $j_1$ to
infect $i$, independently of its siblings
$a,j_2,...,j_{\kappa-1}$. Thus, $p\pi_t$ is the probability of a
sibling of $j_1$ to be infected by $i$, and then reinfect $i$. From
this, $\pi_t\binom{\kappa-1}{k'}\left(p
\pi_t\right)^{k'}\left(1-p\pi_t\right)^{\kappa-1-k'}$ is the
probability that \emph{exactly} $k'$ siblings of $j_1$, in addition
to $j_1$ itself, have succeeded to reinfect $i$. However, after the
first out of these $k'+1$ nodes have reinfected $i$, none of the
other $k'$ nodes can do so. Since all of these nodes are a-priori
equivalent, the probability of $j_1$ to successfully reinfect $i$ is
obtained by dividing by $k'+1$. Finally, the desired probability
$\pi$ is obtained by summing over all possible values of $k'$.

\begin{figure}[h]
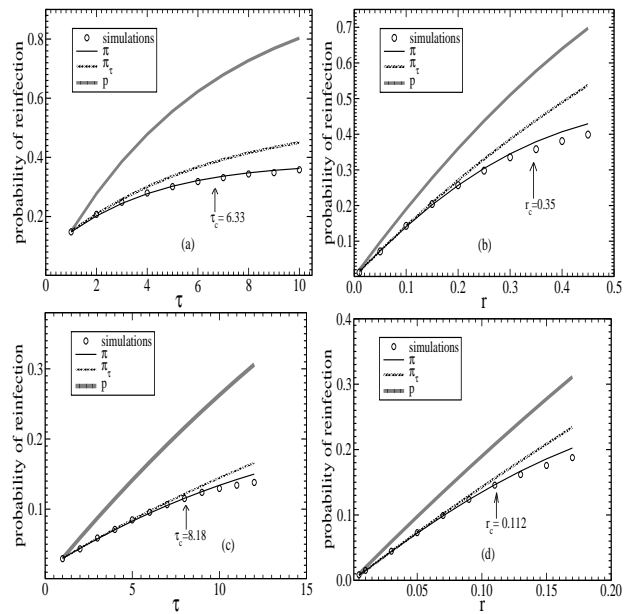

\begin{center}
\epsfig{file=2a.eps,height=4cm,width=4cm}
\epsfig{file=2b.eps,height=4cm,width=4cm}
\epsfig{file=2c.eps,height=4cm,width=4cm}
\epsfig{file=2d.eps,height=4cm,width=4cm}
\end{center}
\caption{Simulation and theory for the reinfection probability.
Three approximations: $p$ (dot-dashed lines),$\pi_t$ (dashed
lines),$\pi$ (solid lines) are compared with simulation results (see
text). Simulation results (circles) were obtained by recording the
probability of a infected node to reinfect its ancestors,
averaging over many network and disease configurations. The
simulations presented are for ER networks but were validated also for SF networks. The epidemic threshold is indicated
with an arrow, and is calculated from Eq.
\eqref{critical_infection_rate}. (a) $\kappa=2.1$ $r=0.15$ and increasing
values of $\tau$. (b) $\kappa=2.1$ $\tau=2$ and increasing values of $r$
(c) $\kappa=5$ and $r=0.03$. (d) $\kappa=5$ and $\tau=2$.} \label{prob_to_reinfect}
\end{figure}

Next, we compare simulation results for the reinfection probability
with the three approximations presented above: (\emph{i}) $p$ - the
reinfection probability assuming it is equal to the probability to
infect any other node (mean field approach); (\emph{ii}) $\pi_t$
- the reinfection probability taking into account only the ``time''
effect (Eq. \eqref{pi_t}); (\emph{iii}) $\pi$ - the reinfection
probability according to Eq. \eqref{formula_pi}. Recall that due to
the assumption explained above, the equation for $\pi$ is expected
to hold only at or below the epidemic threshold. Fig.\ref{prob_to_reinfect} shows that while $\pi_t$ is a better
estimator of the reinfection probability compared to the mean-field
approximation, our Eq. \eqref{formula_pi} is superior to both and
agrees with the simulation results up to the epidemic threshold
(indicated with an arrow).
For increasing values of $\tau$ (Figs. \ref{prob_to_reinfect}(a) and (c))
two opposing effects occur: While $p$ by itself grows with $\tau$,
increasing the probability of reinfection, the growth of $\tau$ also
reinforces the ``time'' and ``neighbors'' effects, thus attenuating
the increase in $\pi$. An increase in $r$ (Figs.
\ref{prob_to_reinfect}(b) and (d)) has a similar outcome. As $r$
increases, $p$ increases. In addition, large values of $r$ imply also
that $j_2,...,j_{\kappa-1},a$ are more probable to infect $i$ before
$j_1$, thus decreasing the total reinfection probability $\pi$. 

Our analysis has so far been general for all random networks with a
prescribed degree distribution $P(k)$. Two specific models are of
special interest. The first is the Erd\H{o}s - R\'{e}nyi (ER) network model
\cite{ER,Bollobas,ER2}, in which all links exist with equal
probability $\phi$, leading to a Poisson degree distribution
$P(k)=e^{-\av{k}}\av{k}^k/k!$ with average degree
$\av{k}=(N-1)\phi$. The ER network model has become a classic model
in random graph theory and was intensively studied in the past few
decades. The other model is that of scale free networks (SF)
\cite{Barabashi}: networks with a broad degree distribution, usually
in the form of a power-law, $P(k) \sim k^{-\gamma}$ with $\gamma>2$.
It was found that many natural networks are scale-free
\cite{Barabashi}.

Using our results, the formulation of the critical infection rate
for the SIS epidemic model on both ER and SF networks is
straightforward. At the threshold,
\begin{equation}
\label{critical_infection_rate} p(r_c)(\kappa-1)+\pi(r_c)=1.
\end{equation}
For ER networks, $\kappa-1=\av{k^2}/\av{k}-1=\av{k}$. Thus, the
equation for the critical infection rate is
$p(r_c)\av{k}+\pi(r_c)=1$. For SF networks, the value of $\kappa-1$
depends on the degree exponent $\gamma$. Theoretically, $\kappa$
diverges for infinite SF networks with $2<\gamma<3$, leading to the
trivial solution $r_c=0$. In practice, even for $\gamma<3$ finite
networks have finite degrees and thus the epidemic threshold is
non-zero \cite{Finite_networks}.

\begin{figure}[h]
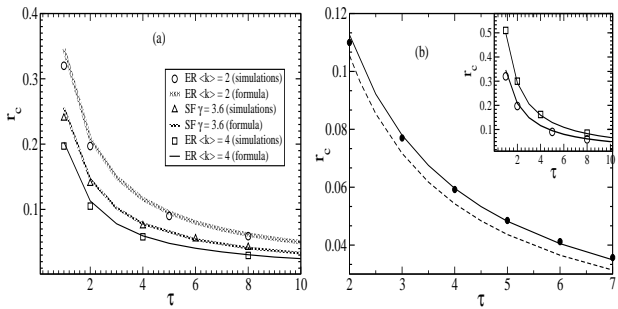

\begin{center}
\epsfig{file=3a.eps,height=4cm,width=4cm}
\epsfig{file=3b.eps,height=4cm,width=4cm}
\end{center}
\caption{Simulation results for the critical infection rate. (a)
Simulation results of $r_c$ vs. $\tau$ for ER networks with
$\av{k}=2,4$ and SF networks with $\gamma=3.6$ (symbols), compared
to theory (lines) as obtained from Eq.
\eqref{critical_infection_rate}. (b) Theoretical values of $r_c$
(Eq. \eqref{critical_infection_rate}; solid line) compared to the
mean-field approach (obtained by setting $\pi=p$ in Eq.
\eqref{critical_infection_rate}; dashed line) and simulation
(circles) for $\av{k}=4$. The inset compares $r_c$ (theory and
simulations) between SIR (circles) and SIS (squares) with
$\av{k}=2$. Theoretical values (lines) were obtained from Eq.
\eqref{rc_sir} for SIR and from Eq. \eqref{critical_infection_rate}
for SIS. The epidemic threshold was calculated using a method of
percolation theory: denote by $S$ the total number of infections;
$r_c$ was identified as $r$ for which $S \sim N$ \cite{Percolation,Percolation3}.}
\label{SIS_criticality}
\end{figure}

Simulation results for the critical infection rate $r_c$ are
presented in Fig. \ref{SIS_criticality}, with the theoretical values
of $r_c$ obtained by numerically solving Eq.
\eqref{critical_infection_rate} for given $\tau$ and $\kappa$. In
Fig. \ref{SIS_criticality}(a) we plot theory and simulations for
$r_c$ vs. $\tau$ for ER and SF networks. In Fig.
\ref{SIS_criticality}(b) we compare our theory, the mean-field
approach, and simulation results. For large values of $\tau$, the
the mean-field approach deviates by up to 10 percent, while our
theory is accurate (see note \cite{Note5}). The inset of Fig.
\ref{SIS_criticality}(b) compares SIR and SIS.

The SIS epidemic model we analyzed has been limited to the case of a fixed recovery time.
This is because, to calculate the epidemic threshold, we
based our theory on the mapping between bond percolation and the SIR
model \cite{Percolation_SIR2,condition_epidemics}. However, it has
recently been shown that this mapping is valid only when the
recovery time is fixed, whereas variable recovery time introduces
correlations not present in the bond percolation model \cite{Kenah}.
Nevertheless, we argue that our reinfection probability $\pi$ is an upper bound 
for the reinfection probability with variable recovery time. This can be shown by simulations as
well as by the following intuitive argument. Consider the simple
case when only the time effect is taken into account and $\pi=\pi_t=[1-(1-r)^{(\tau+1)}]/(2-r)$. 
For a fixed recovery rate, $\tau$ is a random variable, geometrically
distributed with mean $\av{\tau}$. Noting that $\frac{\di^2 \pi}{\di
\tau^2} \leq 0$ for all $\tau$, it follows from Jensen's inequality that
$\av{\pi(\tau)} \leq \pi(\av{\tau})$.

In summary, we analyzed the propagation of diseases on random
networks according to the SIS model. The presence of reinfections in
the model poses a great challenge since it can no longer be mapped
into a simple branching process. We overcame this problem by
calculating the reinfection probability. With this probability, we
could recast the problem back into a branching process with an
effective branching factor that takes into account not only the
descendants of a node, but also its ancestor, weighted by the
probability of reinfection. The condition for an outbreak was then
found using the reinfection probability and percolation arguments.


We thank the European EPIWORK project and the Israel Science
Foundation for financial support and R. Cohen for discussions. S.C.
is supported by the Adams Fellowship Program of the Israel Academy
of Sciences and Humanities.

\end{document}